\documentclass[
  journal=largetwo,
  manuscript=article-type,
  year=2020,
  volume=37,
]{cup-journal}

\usepackage{amsmath}
\usepackage[nopatch]{microtype}
\usepackage{booktabs}

\title{Determination of wind-fed model parameters of High-Mass X-ray Binaries}

\author{Ali Taani}
\affiliation{Physics Department, Faculty of Science, Al-Balqa Applied University, 19117 Salt, Jordan}
\email[Ali Taani]{ali.taani@bau.edu.jo}

\author{Shigeyuki Karino}
\affiliation{Faculty of Science and Engineering, Kyushu Sangyo University, 2-3-1 Matsukadai, Higashi-ku, Fukuoka 813-8503, Japan}

\author{Liming Song}
\affiliation{Key Laboratory of Particle Astrophysics, Institute of High Energy Physics, Chinese Academy of Sciences, Beijing
100049, China}

\author{Chengmin Zhang}
\affiliation{National Astronomical Observatories, Chinese Academy of Sciences, Beijing 100012, China}
\affiliation{School of Physics, University of Chinese Academy of Sciences, Beijing 100049, China}

\author{Sylvain Chaty}
\affiliation{National Astronomical Observatories, Chinese Academy of Sciences, Beijing 100012, China}

\keywords{keyword entry 1, keyword entry 2, keyword entry 3} 

\begin{document}

\begin{abstract}
We have studied several neutron star high-mass X-ray binaries (HMXBs) with super-giant (SG) companions using a wind-fed binary model
associated with the magnetic field. By using the concept of torque balance, the magnetic field parameter determines the mass accretion
rate. This would help us to consider the relationship between wind velocity and mass-loss rate. These parameters significantly improve our
understanding of the accretion mechanism. The wind velocity is critical in determining the X-ray features. This can be used to identify the
ejection process and the stochastic variations in their accretion regimes. However, even in systems with a long orbital period, an accretion
disk can be created when the wind velocity is slow. This will allow the HMXB of both types, SG and Be, to be better characterised by deriving
accurate properties from these binaries. In addition, we have performed segmentation in the parameter space of donors intended for several
SG-HMXB listed in our sample set. The parameter space can be categorised into five regimes, depending on the possibility of disk formation
associated with accretion from the stellar wind. This can give a quantitative clarification of the observed variability and the properties of
these objects. For most of the systems, we show that the derived system parameters are consistent with the assumption that the system is
emitting X-rays through direct accretion. However, there are some sources (LMC X-4, Cen X-3 and OAO1657-415) that are not in the direct
accretion regime, although they share similar donor parameters. This may indicate that these systems are transitioning from a normal wind
accretion phase to partial RLOF regimes. 
\end{abstract}

\section{INTRODUCTION }
\label{sec:intro}

The detection of Cyclotron Resonance Scattering Features (CRSFs) in spectra of many 
accreting neutron stars (NSs) with high magnetic field (B$\geq
10^{12}$ G) provides valuable insights into the physics of emitting
regions and the evolution of these systems. They form due to resonant
scattering processes with electrons, protons, and other ions in the
plane and perpendicular to the magnetic field (Voges et al. 1982;
Wilson et al. 2008; Ye et al. 2020). As a result, the cyclotron line features
provide the only direct estimate of the magnetic field strength of
NSs in X-ray binary systems. In High-Mass X-ray Binaries (HMXBs), a
NS accretes matter from a companion star via stellar wind. The
accreted matter is channeled along field lines of the strong
magnetic field of the NS onto the magnetic poles. X-ray emission
from the NS is produced in regions around the magnetic poles. It is
noteworthy to mention here that most observed cyclotron lines have
been detected above 10 keV and are interpreted as electron features,
with inferred magnetic fields $B\sim 10^{12}$ G (Heindl et al.
2001). The combined effects of poor statistics, photoelectric
absorption and the lack of evidence for a remnant accretion disk
have made these energy sources
elusive. 

Most efforts to calculate theoretical cyclotron lines have been
performed in a line-forming region with a constant temperature and
density of an electron-proton plasma permeated by a uniform magnetic
field (Wheaton et al. 1979; Orlandini et al. 1999; Yamamoto et al.
2011; Ye et al. 2019). 
 Nishimura (2005) calculated cyclotron lines assuming a
strong variation in field strength with distance from an emission
region. However, no model generating such high flux and high
temperature at a layer deeper than absorbing heavy atoms has been
proposed.

According to recent studies, several pulsars show changes in
luminosity dependence in the cyclotron resonance energy. The first
aim of this paper is to derive magnetic field strengths, which is
crucial for these systems, 
and obtain clues about
the evolution of HMXBs, which can be understood in terms of the
conservative evolution of normal massive binary systems.

The second aim of this study is to derive unknown parameters of
HMXBs without uncertainty in the strength of the NS magnetic field.
With robust data on the NS magnetic field, combined with spin period
($P_{\rm{spin}}$) and orbital period ($P_{\rm{orb}}$), we can fix
several hitherto-unknown parameters, such as wind velocity and wind
mass loss rate. These parameters influence significantly the model
of wind-fed binary systems and can constrain the effects of binary
evolution (Taani \&  Khasawneh 2017; Dai et al. 2017; Taani et al. 2019a,b;
Karino et al. 2019; El Mellah et al. 2019a,b; Karino 2020). From
this standpoint, with observations of NS magnetic fields, we could
constrain the end products of HMXBs, such as a NS-NS merger, which
is considered to be one of the most powerful gravitational wave
sources and also the most probable site for heavy element creation
(Taani 2015; Postnov \& Yungelson 20016; Haniewicz et al. 2020).

In the next section, we introduce the recent results of NS magnetic
field given by CRSF observations. In Section 3, we discuss the
method to obtain hitherto-unknown binary parameters from robust data
on the NS magnetic field in SG-HMXBs. In Section 4, we discuss our
findings. The last section is devoted to conclusions.
 
\section{Cyclotron lines}

Since the physical conditions are expected to vary over the emission
region, the X-ray spectrum is expected to change with the viewing
angle and therefore with pulse phase. This variation can be because
during one rotation phase different parts of the surface are exposed
and also due to change in local field structure due to accretion
dynamics, e.g. change in accretion rate, as is seen for sources like
Her X-1 and V0332+53. The difference in time scales of variation for
E$_{cyc}$ and luminosity will allow researchers to distinguish
between
these two distinct cases. 
(Nagase et al. 1991; Wilson et al. 2008). In this work, we have
selected 11 persistent sources with SG companions known to have at
least one cyclotron line (see Table 1). Here, our analysis of all
pointing observations provides an opportunity to infer the values of
magnetic field strength according to their spectra. However, the
gravitational redshift \emph{z} that at the NS surface is
approximately

\begin{equation}
z\simeq\frac{1}{\sqrt{1-\frac{2GM_{\rm{NS}}}{R_{\rm{NS}}c^{2}}}}-1\simeq0.3
\end{equation}


where 
$M_{\rm{NS}}, R_{\rm{NS}}$ and $c$ are the NS mass, radius and speed
of light, respectively. Assuming canonical values for $M_{\rm{NS}}$
and $R_{\rm{NS}}$ of 1.4 $\rm{M}_{\odot}$ and 10 km, thus \emph{z} =
0.3. As such, the line energy of the fundamental cyclotron line is
related to the magnetic field strength by the equation
\begin{equation}
\begin{array}{rcl}
       E_{\rm cyc}=11.6B_{12}(1+z)^{-1}.
\end{array}
\end{equation}

Here \emph{B}$_{12}$ is the magnetic field strength in units of
10$^{12}$ G, and the higher harmonics have an energy \emph{n}
times the fundamental energy $E_{\rm{cyc}}$. 

In general, the magnetic field of NSs spans a range from $10^{8}
\rm{G}$ or less (LMXBs) to $10^{15} \rm{G}$ (magnetars); this
concentration seems rather odd. Additionally, the magnetic field
shows no correlation with the spin period of NSs and orbital period
of the binary systems (see Fig. 1) This result disagrees with the
dependence of B-field on the spin period in Be-systems shown by
previous studies (Corbet 1986,2004; van den Heuvel 2009; Mardini et
al. 2019a,b). Though, of course, we need to consider some
observational biases: too strong of a magnetic field prevents
accretion onto the NS and we could not observe such systems as
bright X-ray sources (Taani 2016). Besides such possibility of biases, this
concentration around $B \approx 10^{12} \rm{G}$ will draw a lot of
interest and promote further studies on the NS magnetic field (Taani et al. 2019a,b). The
fundamental energy covers a wide range, starting at 10 keV for Swift
J1626.6-5156 (DeCesar et al. 2009) to 100 keV for LMC X-4 (La
Barbera  et al. 2001).


%
The strong energy variation of the cyclotron lines (for example, in
V0332+53, GRO J1008-57 and GX301-2) can be used to argue that during
different phases of the X-ray pulses, regions with different
magnetic fields are observed.

It is noteworthy to mention here that 4U 0115+634 is one of the
pulsars whose CRSFs have been studied in great detail (see, e.g.,
Wheaton et al. 1979; Nagase et al. 1991; Nishimura 2005). In
previous outbursts, CRSFs have been detected up to the fifth
harmonic (Heindl et al. 2003; Ferrigno et al. 2011). This high
number of detected CRSFs in 4U 0115+634 makes this system an
outstanding laboratory to study the physics of cyclotron lines in
X-ray pulsars.


\begin{figure}
\includegraphics[width=10cm]{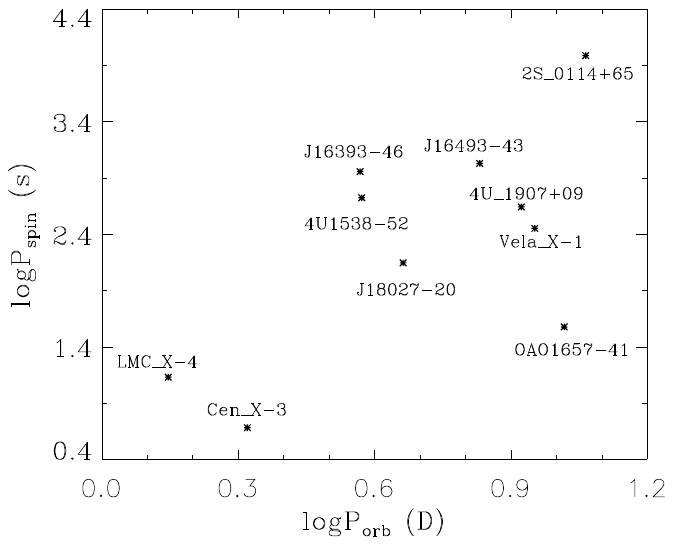}
\caption{The Corbet diagram of SG-HMXBs.}
\end{figure}

\newcounter{rit} 
\newcommand{\rit}{\refstepcounter{rit}\therit}

\begin{table*}
\caption{List of some observational parameters of all known
persistent sources with supergiant companions through the cyclotron
resonant scattering features} \label{O-I}
\setlength{\tabcolsep}{2pt}
\renewcommand{\arraystretch}{0.1}
\begin{tabular}{lcccccl}
\hline \hline \noalign{\smallskip}
 \multicolumn{1}{c}{Object} &
\multicolumn{1}{c}{$P_{\rm spin}$} & \multicolumn{1}{c}{$P_{\rm
orbit}$} & \multicolumn{1}{c}{$E_{\rm cyc}$} &
\multicolumn{1}{c}{Ref.} \\
\multicolumn{1}{c}{}& \multicolumn{1}{c}{(s)} & \multicolumn{1}{c}{(d)} &
\multicolumn{1}{c}{(keV)} &
\multicolumn{1}{c}{}\\
\hline \noalign{\smallskip}
\\ 
4U 1907$+$09 & 439 & 8.37 & 18.8$\pm$0.4  &              \ref{1998Cusumano}, 
              \ref{2002Coburn}, \ref{2010Rivers}\\
4U 1538$-$52 & 529 & 3.73
             & 21.4$^{+0.9}_{-2.4}$  &
             \ref{2002Coburn}, \ref{1990Clark}, 
              \ref{2001Robba},  \ref{2009Rodes-Roca} \\
Vela X-1 & 283 & 8.96 & 27$^{+0.5}_{-1.1}$ &
             \ref{1996Kretschmar}, \ref{1999Makishima}, \ref{2002Kreykenbohm}, \ref{2007Schanne}\\
Cen X-3 & 4.8 & 2.09 & 30.4$^{+0.3}_{-0.4}$ &        \ref{2002Coburn}, \ref{1999Makishima}, \ref{1998Santangelo}  \\
LMC X-4 & 13.5 & 1.4 & 100$\pm$2.1 &
         \ref{1999Makishima}, \ref{2001LaBarbera} \\

 OAO1657-415 & 37.7& 10.4& 36&
 \ref{1999Orlandini}, \ref{2010Denis}, \ref{2011Pottschmidt}\\
J16493-4348$^{\dag}$ & 1069 & 6.78 & 33$\pm$4  &  \ref{2010Nespoli}, \ref{2011D'Ai}  \\
4U 1700-377 & --&3.4& 37&    \ref{2011Pottschmidt}, \ref{1999Reynolds}\\
2S 0114+65& 9700& 11.6& 22&   
\ref{2005Bonning}, \ref{2006denHartog}\\
J16393-4643&904&4.2&29.3$^{+1.1}_{-1.3}$& \ref{2016Bodaghee}\\
IGR J18027-201&140&4.6&23& \ref{2017Lutovinov}\\
\hline \noalign{\smallskip}
\end{tabular}

References.-- These references are to period measurements in the literature.
  Some have errors originating from applied analysis, designated
with a dagger, or from the supplied data, designated with an
asterisk. (\rit\label{1998Cusumano}) Cusumano et al. 1998;
 (\rit\label{2002Coburn}) Coburn et al. 2002;
 (\rit\label{2010Rivers}) Rivers et al. 2010;
 (\rit\label{1990Clark}) Clark et al. 1990;
 (\rit\label{2009Rodes-Roca}) Rodes-Roca et al 2009;
 (\rit\label{2001Robba}) Robba et al. 2001;
 (\rit\label{1996Kretschmar}) Kretschmar et al. 2005;
  (\rit\label{1999Makishima}) Makishima et al. 1999;
  (\rit\label{2002Kreykenbohm}) Kreykenbohm et al. 2002
 (\rit\label{2007Schanne}) Schanne  et al. 2007
   (\rit\label{1998Santangelo}) Santangelo et al. 1998;
(\rit\label{2001LaBarbera}) Barbera et al. 2001;
(\rit\label{1999Orlandini}) Orlandini et al. 1999;
(\rit\label{2010Denis}) Denis et al. 2010;
(\rit\label{2011Pottschmidt}) {Pottschmidt}, S et al. 2011
  (\rit\label{2010Nespoli}) {Nespoli} et al. 2010
  (\rit\label{2011D'Ai}) {D'Ai} et al. 2011
(\rit\label{1999Reynolds}) {Reynolds} et al. 1999
(\rit\label{2005Bonning}) {Bonning} et al. 2005
(\rit\label{2006denHartog}) {denHartog} et al. 2006
(\rit\label{2016Bodaghee}) {Bodaghee} et al. 2016.~
(\rit\label{2017Lutovinov}) {Lutovinov} et al. 2017.~
\end{table*}

\section{Investigating wind parameters in SG HMXB systems}

Under the assumption that the spin period of a NS is nearly in
equilibrium, its magnetic field strength can be estimated by
\begin{equation}
B_{\rm{NS}} = 2.184 \times 10^{12} \rm{G} \times \zeta^{1/2}
\left( \frac{\dot{M}}{10^{18} \rm{g\, s}^{-1}} \right)^{1/2}
\left( \frac{P_{\rm{s}}}{1 \rm{s}} \right)^{7/6} ,
\label{eq:B}
\end{equation}
assuming that the mass of the NS  is 1.4 $M_\odot $ and its radius
is 10~km (Ghosh \& Lamb 1979; Campana et al. 2002; Tsygankov et al.
2016). The parameter $\zeta$ is the
ratio of accretion velocity to the free-fall velocity, and hereafter
we fix this value as 0.5.

The mass accretion rate $\dot{M}$ can be obtained as the following,
if we assume the Hoyle-Lyttleton accretion scenario (Karino
et al. 2019).
\begin{equation}
\dot{M} = \dot{m}_{\rm{w}} \times \frac{G^2 M_{\rm{NS}}^2 }{a^2 v_{\rm{w}} v_{\rm{rel}}^3} ,
\end{equation}
%
where $\dot{m}_{\rm{w}}$ is the wind mass loss rate from the donor.
Here, orbital radius $a$ can be obtained from the orbital period of
the system. The X-ray luminosity is related to the mass accretion
rate as follows
\begin{equation}
L_{X} \simeq \frac{G M_{\rm{NS}} \dot{M}}{R_{\rm{NS}}} .
\label{eq:LX}
\end{equation}
Thanks to this relation, we can deduce the mass accretion rate from the observed X-ray luminosity.
The relative velocity of the wind to the NS is
\begin{equation}
v_{\rm{rel}} = \left( v_{\rm{orb}}^2 + v_{\rm{w}}^2 \right)^{1/2}
\end{equation}
and the velocity of the line-driven wind is  usually
prescribed in the model by so-called $\beta$-law . 
\begin{equation}
v_{\rm{w}} = v_{\infty} \left( 1 - \frac{R_{\rm{d}}}{a} \right)^{\beta}
\label{eq:vwind}
\end{equation}

In this study, we assume $\beta$, which is a free input parameter,
to be $\beta = 1$ (Puls et al. 2008). $v_{\infty}$ denotes the
terminal velocity of the wind.

The wind parameters such as $\dot{m}_{\rm{w}}$ and $v_{\infty}$
contain large uncertainties. Combining above equations with CRSF
data introduced in the previous section, however, a single
relationship between $\dot{m}_{\rm{w}}$ and $v_{\infty}$ could be
obtained for the canonical NS parameters. From Eqs.~(\ref{eq:B}) to
(\ref{eq:vwind}), we get
\begin{equation}
v_{\rm{w}}^{2} = - v_{\rm{orb}}^{2} \pm \sqrt{ \frac{G M_{\rm{NS}}^{2}}{\pi a^2 \dot{M}}
 \dot{m}_{\rm{w}}} .
\label{eq:XXX}
\end{equation}
If the luminosity of the NS is known, the mass accretion rate $\dot{M}$ could be derived
from Eq.~(\ref{eq:LX}).
$\emph{a}$ is the orbital semi-major axis and it is obtained from the orbital period if the mass
of the donor is known.
In Table~\ref{tab:2}, we show the values of donor mass and donor radius,
which has appeared in Eq.~(\ref{eq:vwind}).

If we choose the mass loss rate from the donor $\dot{m}_{\rm{w}}$ as
a fundamental variable, then Eq.~(\ref{eq:XXX}) becomes a
biquadratic equation and has four solutions. Two of them are
physically nonsense, so we consider the other two solutions since
they have a clear correlation between wind velocity and the mass
loss rate of donors in SG-HMXBs.
These solutions of the wind velocity are shown in Figs. 2 - 4 by
solid curves, as functions of mass loss rate in SG-HMXBs. In these
figures, the wind parameter given by the frequently used wind model
by Vink, de Koter, \& Lamers (2001) are shown at the same time. In Vink, de Koter, \& Lamers (2001), the
approximated wind mass loss rates from SG stars are given by the
complex functions of mass, radius (escape velocity), luminosity
(effective temperature) of SG stars. To derive the mass loss rate of
the HMXB donors, we use the mass and radius data introduced in
previous papers (Chaty et al. 2008; Cusumano et al. 2010; Rawls et al.
2011;Mason et al. 2012; Falanga et al. 2015; Reig et al. 2016). From
these data, we derive the effective temperature and corresponding
luminosity using approximated stellar evolution scheme. We compile
Eqs. (1) to (30) in  and find the evolution stage of
each donors in SG-HMXBs listed in Table~\ref{tab:2}. Then, we
derive the effective temperature at this evolutionary stage.

\begin{table*}\caption{List of derived parameters for SG HMXBs}
\setlength{\tabcolsep}{5pt}
\renewcommand{\arraystretch}{1}
  \begin{tabular}{|l|c|c|c|c|c|c|c|} \hline
    name & $B_{\rm{NS}}$ [$10^{12}$G] & $M_{\rm{d}}$ [$M_{\odot}$] & $R_{\rm{d}}$ [$R_{\odot}$]
        & $\log (L_{\rm{d}} / L_{\odot})$ & $T_{\rm{eff}}$ [$10^4 \rm{K}$]&  $v_{\infty}$ [$10^7 \rm{cm \, s}^{-1}$]
            & $\dot{m}_{\rm{w}}$ [$10^{-8} M_{\odot} \rm{s}^{-1}$] \\ \hline \hline
    LMC X-4 &  11.2 & 15.0 & 7.7 & 4.50 & 2.78 & 22.4 & 1.10 \\ \hline
    Cen X-3 & 3.4 & 22.1 & 12.6 & 4.92 & 2.77 & 21.3 & 5.37 \\ \hline
    J16393 &  3.3 & 20 & 13 & 4.82 & 2.57 & 9.96 & 37.2 \\ \hline
    4U1538 &  2.4 & 14.1 & 12.5 & 4.67 & 2.40 & 8.53 & 25.7 \\ \hline
    J18027 &  2.6 & 21 & 19 & 4.90 & 2.22 & 8.44 & 44.7 \\ \hline
    J16493 &  3.7 & 47 & 32 & 5.64 & 2.62 & 19.5 & 776 \\ \hline
    4U1907 &  2.1 & 27.8 & 22.1 & 5.21 & 2.47& 9.01 & 166 \\ \hline
    Vela X-1  & 6 & 24.0 & 31.8 & 5.21 & 2.06 & 6.98 & 3.72 \\ \hline
    OAO1657 &  4 & 14.3 & 24.8 & 4.72 & 1.76 & 6.10 & 22.9 \\ \hline
    2S0114 &  2.5 & 16 & 37 & 4.85 & 1.55 & 5.28 & 33.9 \\ \hline
\end{tabular}
\label{tab:2}
\\
The data on mass and radius of NS are taken from
(Falanga et al. 2015; Reig et al. 2016; Chaty et al. 2008;
Rawls Reig et al. 2011; Cusumano et al.2010; Mason et al.2012) . The luminosity and
effective temperature are computed by their approximated stellar
evolution track given by Hurley  et al. (2001).

From these donor data, the terminal velocity of the wind and the wind mass loss rate from SG
stars are derived by polynomial approximation given by Vink  et al. (2001).
\end{table*}

The wind parameters ($v_{\infty}$ and $\dot{m}_{\rm{w}}$) we obtained are reported in Table 2.
From these results, we could confirm that terminal velocities of
the wind from donors in SG HMXBs are rather slow. In seven systems,
the terminal velocities of donors dip from the typical wind velocity
of galactic single SG stars ($1,000-2,000\, \rm{km \, s}^{-1}$). This
result is consistent with recent result given by
Gim\'{e}nez-Garc\'{\i}a et al. (2016) who argued that the stellar
wind of donors in persistent HMXBs is systematically slow\footnote{On the other hand, the wind velocity in SFXT systems
seems very fast ($v_{\infty} = 1,500 \rm{km \, s}^{-1}$).}.
For instance, from recent observations, it is suggested that the wind velocity in persistent SG HMXBs, Vela X-1,
is relatively slow ($v_{\infty} = 700 \rm{km \, s}^{-1}$).
In our samples, even for systems with higher $v_{\infty}$, the wind velocities at the NS positions are typically
$v_{\rm{w}} \approx 500 \rm{km \, s}^{-1}$, and still show very slow wind.

It is broadly considered that in wind-fed HMXBs, the wind matter is
captured by the NS magnetic field at a certain radius, and
transported onto the polar regions of the NS. In this process,
around the polar region, the accretion column is formed and the
potential energy of the accretion matter is converted into strong
X-ray radiation.

However, it is believed that, when the NS (and consequently NS
magnetic field lines) rotates rapidly, the accretion matter cannot
fall onto the NS surface and in some conditions it could be expelled
out (Pfahl et al. 2002; Podsiadlowski et al. 2004). This rotational
inhibition of the accretion matter is called the propeller effect
(see Reig \& Zezas 2018). The propeller / accretion limit could be
defined by three typical radii (accretion radius $r_{\rm{a}}$,
magnetic radius $r_{\rm{m}}$ and corotation radius $r_{\rm{co}}$).
The propeller regime is defined when the accretion radius of the
disk is larger than the magnetic radius. In contrast, in the
supersonic inhibition regime, the magnetic radius is larger than the
accretion radius and corotation radius, thus it rotates more slowly
than inner regions of the disk.

The parameter space could divided into five accretion regimes based
on their magnitude relation (Stella, White, \& Rosner 1986;
Bozzo, Falanga, \& Stella 2008). That is, the
parameter space can be categorized into
\begin{itemize}
\item (A) supersonic inhibition regime ($r_{\rm{m}} > r_{\rm{a}}, r_{\rm{co}}$),
\item (B) subsonic inhibition regime ($r_{\rm{co}} > r_{\rm{m}} > r_{\rm{a}}$),
\item (C) supersonic propeller regime ($r_{\rm{a}} > r_{\rm{m}} > r_{\rm{co}}$),
\item (D) subsonic propeller regime ($r_{\rm{co}} , r_{\rm{a}} > r_{\rm{m}}$ , $\dot{M} < \dot{M}_{\rm{c}}$),
\item (E) direct accretion regime.
\end{itemize}
Here, $\dot{M}_{\rm{c}}$ denotes the critical limit where radiative cooling starts working (see Bozzo et al. 2008).

In figures shown below (Figs. 2 - 4), the different accretion
regimes (A) to (E) are divided by dashed lines. The shaded region
denotes the direct accretion regime such that only systems in this
region can be observed as a persistent HMXB. The efficiency of the
propeller depends weakly on the magnetic moment of the star
(Ustyugova et al. 2006). Since if the angular velocity of the star
is larger, then the efficiency of the propeller becomes higher
(Tsygankov et al. 2016). In contrast, in the inhibition regime the
accreting matter will be prevented by the magnetic gate due to being
gravitationally focused toward the NS and thus the centrifugal gate
also propels away material along the magnetospheric boundary of the
NS (Bozzo et al. 2016). In addition, the subsonic propeller regime
becomes clearer as the strength of the propeller increases. As the
strength increases there is a sharp decrease in the accretion rate
to the star.
In these figures, the solid curves represent the theoretical
relations between $\dot{m}_{\rm{w}}$ and $v_{\rm{w}}$ given by
Eq.~(\ref{eq:XXX}) with the CRSF data. We show that when these
curves come into the direct accretion region (shaded region) created
in the figures, the systems with corresponding parameters can lead to X-ray emission.

\begin{figure*}
\includegraphics[width=7cm]{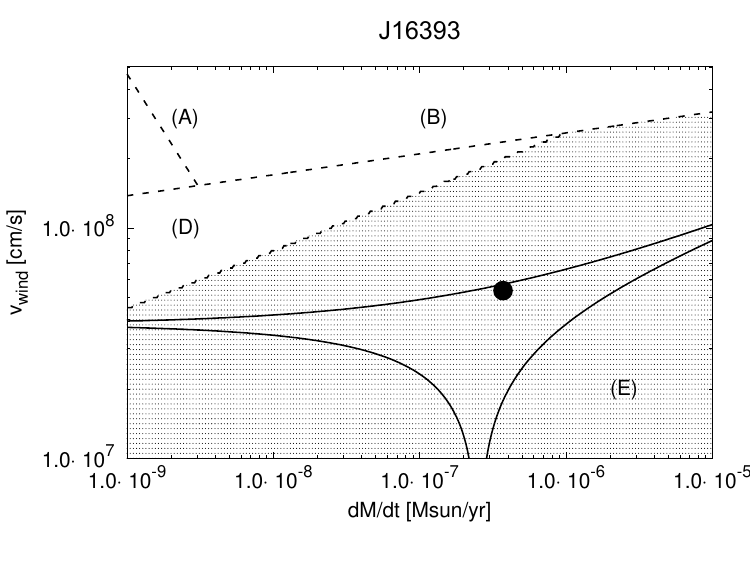}
\includegraphics[width=7cm]{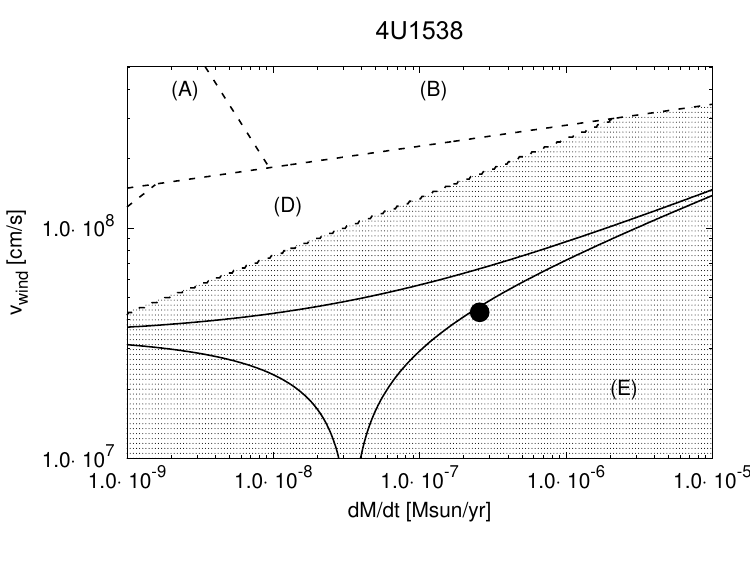} \\
\includegraphics[width=7cm]{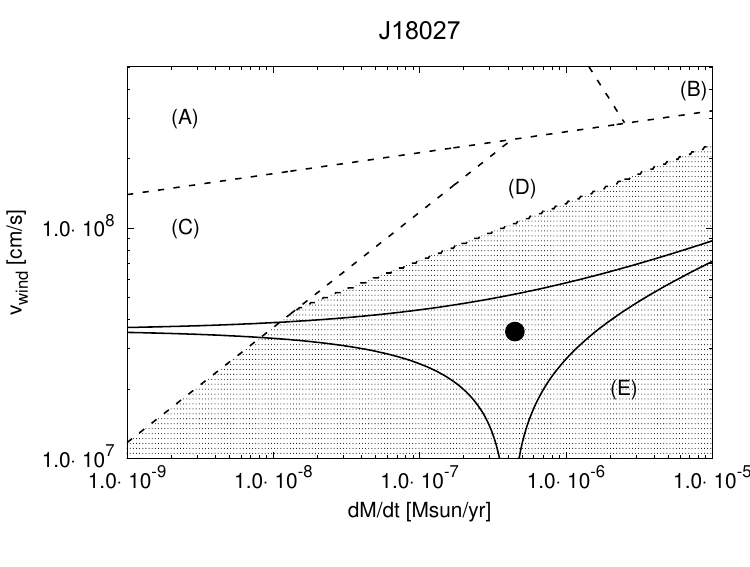}
\includegraphics[width=7cm]{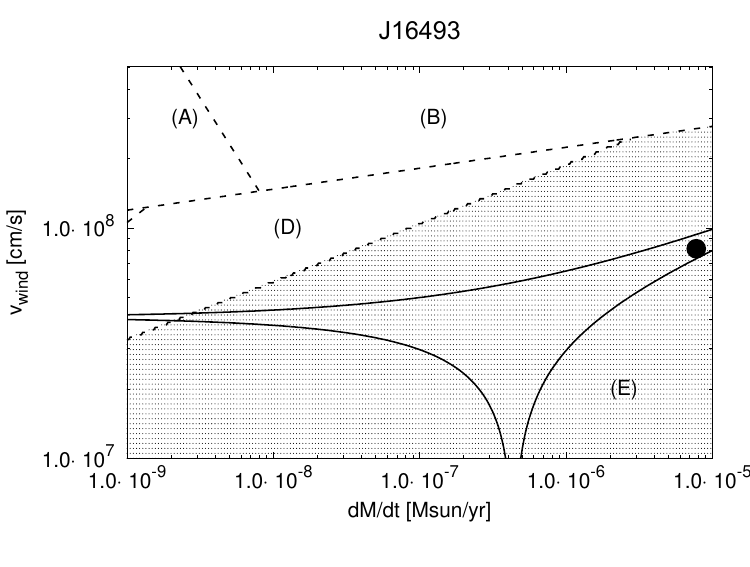} \\

   \caption{Plots of the wind velocity vs mass-loss rate, in various accretion regimes. The position of each source is shown by a black filled circle. As shown clearly, the parameter
space can be categorized into (A) supersonic inhibition regime, (B)
subsonic inhibition regime, (C) supersonic propeller regime, (D)
subsonic propeller regime, and (E) direct accretion regime indicated
by the shaded region, and the solution of the wind equation is
represented by solid curves. } \label{fig:XX}
\end{figure*}

\begin{figure*}
\includegraphics[width=7cm]{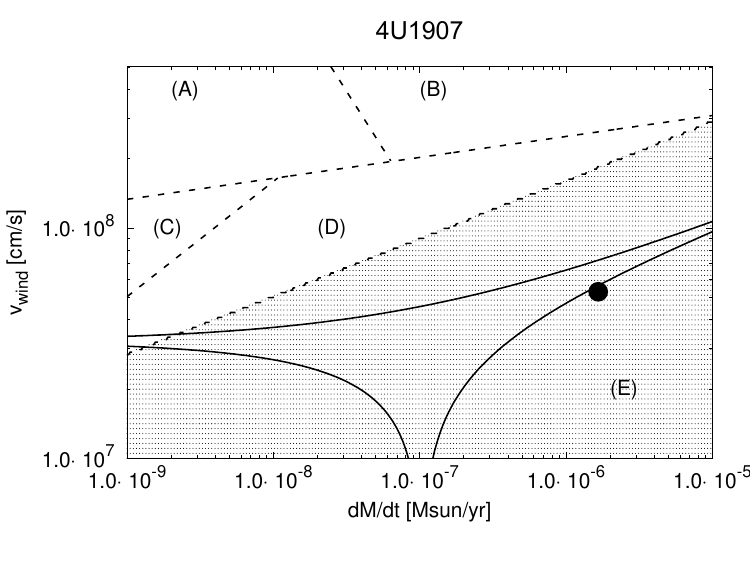}
\includegraphics[width=7cm]{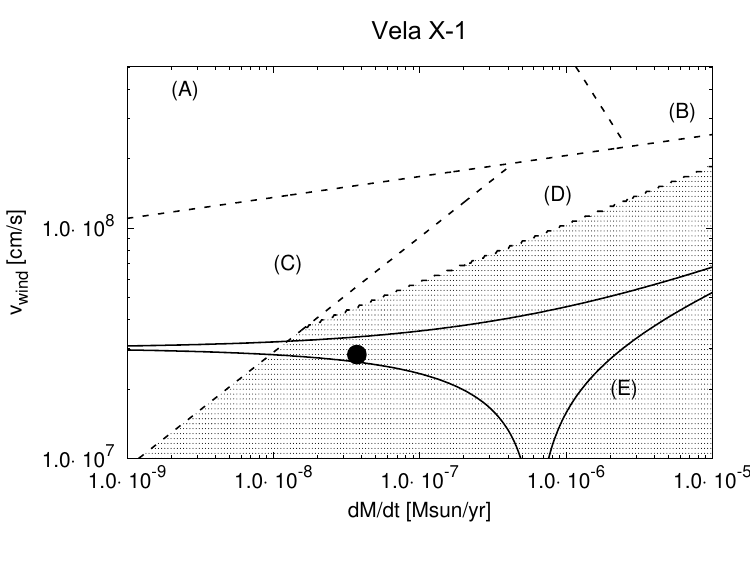} \\
\includegraphics[width=7cm]{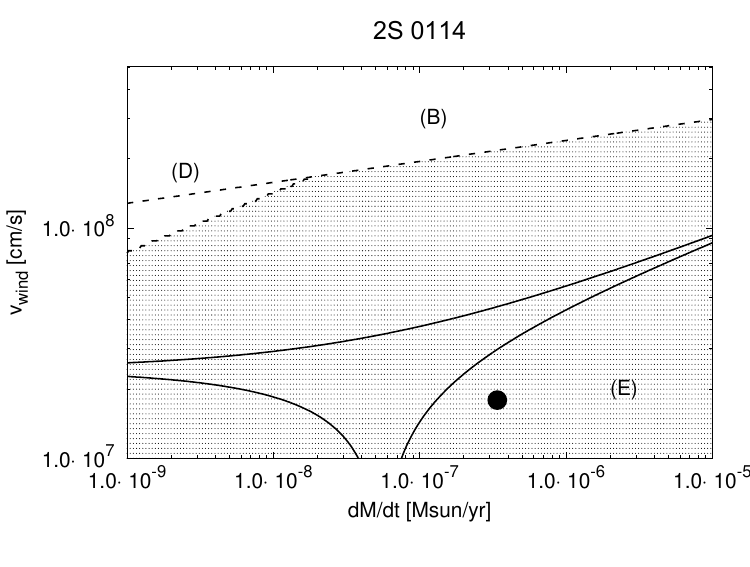}

   \caption{The same as previous, for the wind mass loss-rate,
   accretion mass-loss rate and accretion regimes. The position of each source is shown by a black filled circle.}
\label{fig:YY}
\end{figure*}

\begin{figure*}
\includegraphics[width=7cm]{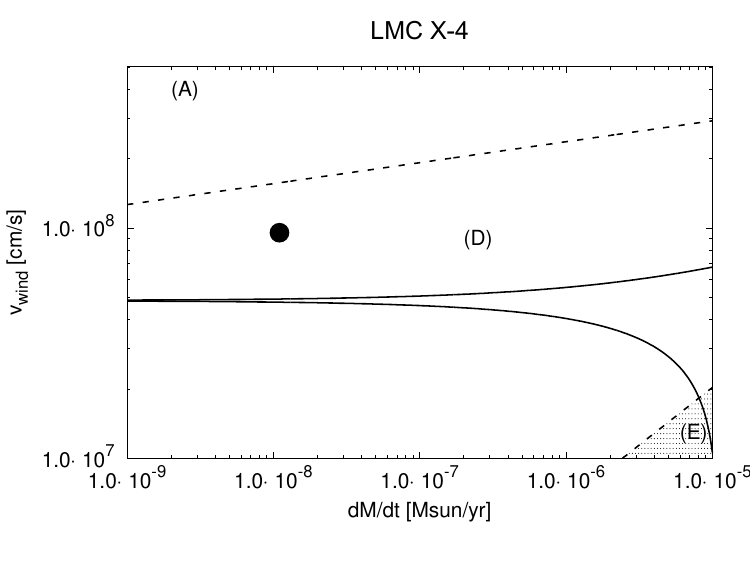}
\includegraphics[width=7cm]{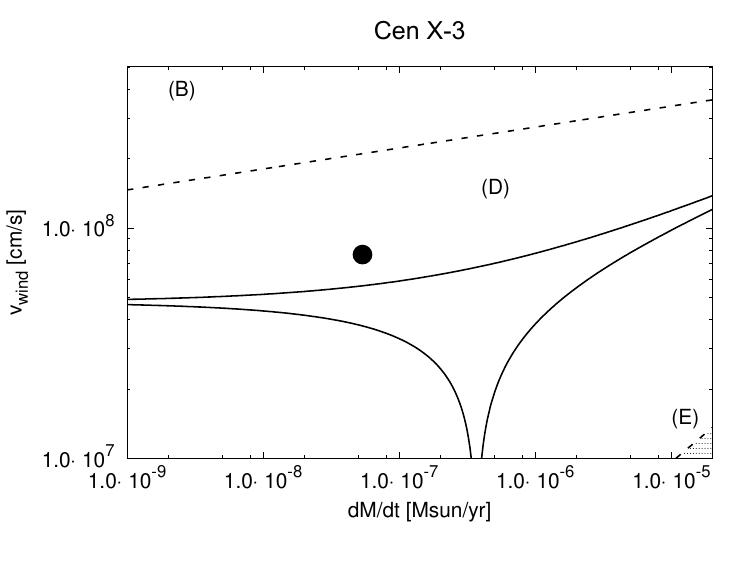} \\
\includegraphics[width=7cm]{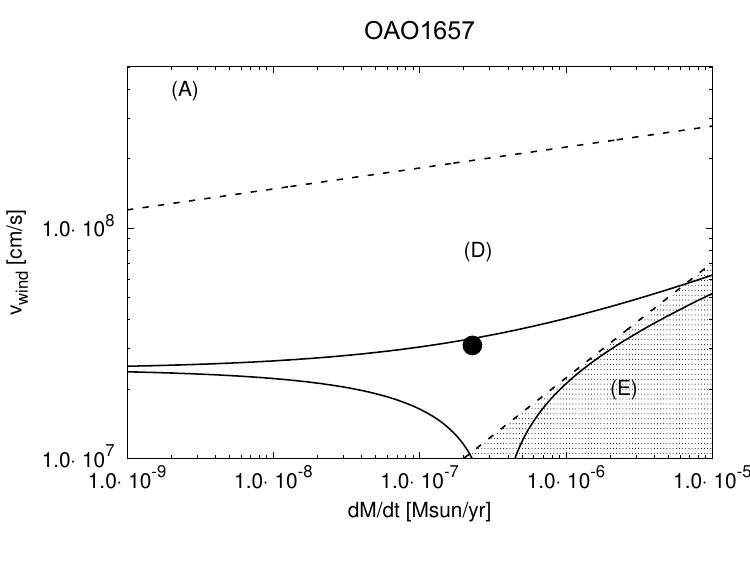}
   \caption{The same initial conditions in terms of the stellar wind parameters as in Figs. 2-3, but these sources show different
   features and different distributions.}
\label{fig:YY1}
\end{figure*}


\section{Discussion}

\subsection{Wind parameters}

In Figures 2-4, we show the direct accretion region where the
systems with corresponding parameters can emit strong X-rays. In the
same figures, we plot the wind parameters given by the standard wind
model combined with the stellar evolution track; furthermore we show
the theoretical relations between the $\dot{m}_{\rm{w}}$ and
$v_{\rm{w}}$ given by Eq.~(\ref{eq:XXX}) with the CRSF data. The
position of each source is based on the results reported in Table 2.
In systems shown in Figures 2 and 3, these plots show good
consistency: the plots are located in the direct accretion region
(shaded region) and roughly follow the theoretical curves (obtained
with CRSF data). These result partly explain the slow wind
tendencies in SG-HMXBs. Namely, when the wind velocity becomes too
high, the wind plots might go outside of the accretion regime from
the upper boundary of the shaded region. For typical mass loss rate
in SG stars (say, $10^{-7} \rm{M}_{\odot} \rm{yr}^{-1}$), the upper
bound of the accretion regime is $\approx 800 \rm{km \, s}^{-1}$.
Then, the systems with fast-wind donors cannot be observed as bright
persistent X-ray sources.

On the other hand, our model cannot be applied for three binaries,
such as LMC X-4, Cen X-3 and OAO1657 (see Fig. 4) although they
share similar donor parameters. Since in systems with shortest
orbital periods (LMC X-4 and Cen X-3) the Roche-lobe filling factors
approach quite near to 1, their accretion mode may not be typical
wind accretion any more. Their accretion mode could enter in the
regime of RLOF, or quasi RLOF (Shakura et al. 2012; Shakura et al.
2013). In this case, it is little wonder that we cannot obtain
consistent wind parameters for these sources. Additionally, OAO1657
shows an inconsistent parameter set. As a result, the Roche lobe
filling factor is much smaller and RLOF cannot be realized. On the
other hand, it is suggested that the donor in this system is a
Wolf-Rayet star (Mason et al. 2009; Mason et al. 2012). In this
case, it might be ill-adopted to the standard wind model for typical
SG stars. It is also argued that the system parameters of OAO1657
cannot be reproduced with standard binary evolution theory: a lot of
mysteries remained in the understanding of this system (Jenke et al.
2012; Walter et al. 2015).



\subsection{NS magnetic field}

The question of where exactly the magnetic field is
measured still remains unanswered.
This depends on the accretion geometry and flow and other mechanisms
(Wei et al. 2010; Coburn et al. 2002; Kreykenbohm et al. 2005).
Hence, their line profiles reflect the geometrical and physical
properties of the accretion column near the magnetic poles of the
NS, and therefore constitute a diagnostic tool for accessing the
physics of accretion.



It is noteworthy to mention here that the NS magnetic fields  in
Table 2 are surprisingly concentrated in a narrow range $\sim
10^{12} \rm{G}$. Despite the fact that their physical properties, in
particulate their energy band (10 - 100 keV) which govern the
evolution, are different, they strongly depend on the assumed
parameters, and these parameters dominate their evolutionary stages.
Thus, the magnetic field itself is of fundamental significance to
having a thorough insight into the physics of the emitting region
structure, and could also be imperative to assisting us in improving
our understanding of binary evolution. Otherwise, the implementation
of known stellar evolution and observational statistics in
population synthesis codes will remain a major issue in our
understanding of the processes occurring in compact binaries or in
the treatment of selection effects (Postnov \& Yungelson 2006).
However, our results shown in Figs. 2 - 4 clearly demonstrate the
variety of SG-HMXBs based on the different types of interactions
between the wind mass loss rate and the three NS radii (accretion
radius, magnetic radius and corotation radius). This diversity of
X-ray binary systems is important in principle, and could be used to
demonstrate the properties of wind-fed systems such as SG-HMXBs, and
the parameters entirely control their evolution, since the binaries
with compact remnants are primary potential GW sources.

Finally, 1A 0114+650 is a unique source  with unusual properties
(very slow rotation period, relatively low X-ray luminosity and a
super-orbital periodicity of 30.7 days (see Farrell et al., 2006)),
exhibiting properties consistent with both Be and SG X-ray binaries
(Walter et al. 2015). This source suggests that it evolves on a time
scale of several years (Wang 2010), or may be an accreting magnetar
candidate (Sanjurjo-Ferrrin et al. 2017; Tong \& Wang 2018).
Possible signature of a transient disk was also found (Hu et al.
2017).

\section{Summary and Conclusions}



The following conclusions and implications are
obtained:

We have determined new physical quantities for several
HMXBs with supergiant companions through their cyclotron
lines. These parameters are: the terminal velocity of the wind, mass-loss rate of the donor, magnetic field, effective temperature and corresponding luminosity. Furthermore, for all systems, our analysis (direct accretion condition shown by shaded region, and wind equation solution shown by solid curves) indicates that the wind velocity must be systematically slow.


By adopting the accretion regime model by  Bozzo et al. (2008), we
have explored the parameter space  in several regimes to support the
intrinsic variabilities of  mass accretion rate and wind velocity.
This may allow us to study an evolutionary path for several
SG-HMXBs in these diagrams. Different regimes are sufficient to distinguish the bright X-ray sources spatially, and the magnetic field-wind velocity can be probed. As a result, persistent SG HMXBs
within the shaded region can be observed through the direct
accretion regime. This interpretation is  predicated on its emission of
accretion in high-energy X-rays.

%

It is seen that the wind velocity causes a significant effect on the results of their X-ray features and it could be used to determine the ejection mechanism.
Consequently, when the wind velocity is slow, the accretion disk is often formed even in systems with large orbital period.
This will allow to better characterize the HMXB of both types, SG and Be, hosting NS, by deriving accurate properties of those compact binaries.


From the updated measurement of HMXB cyclotron lines, the derived
magnetic fields given by CRSF data are all concentrated around $\sim
10^{12} \rm{G}$. However, the fundamental energy during X-ray
observation, spin and other physical parameters property diverges
and varies. The existence of a high magnetic field has the potential
to regulate their formation and evolution.



The accretion mechanism for the fast spinning NSs (P$_{spin}\leq40$ s) with a short orbital period (P$_{orb}\leq10$ d),  like in LMC X-4, Cen X-3 and OAO1657 (see Fig. 4) can not be constrained by our model. In LMC X-4 and Cen X-3, these two binary systems are extremely tight
systems. Thus, the accretion mechanism can therefore not be approximated by
spherical wind, because in such tight systems the concentrated
asymmetric wind or RLOF accretion should be considered. Although OAO 1657 is a further evolved star with a long orbit, the donor of this system can be detected throughout its evolution as a Wolf-Rayet star with a stellar wind mass-loss rate.

Finally, we note that currently  CRSFs data available  are not
sufficiently accurate or numerous to allow for precise analysis. One
would hope that the results of this work will be improved with data
from Suzaku, \emph{INTEGRAL}, $\emph{eRosita}$ and $\emph{HXMT}$,
which can provide significant increase in the observational
sensitivity of a few cyclotron sources.





\textbf{acknowledgements}:
We are grateful to A. D'Ai, E. Nespoli, O. Nishimura, M. Orlandini,
K. Pottschmidt, R. Rothschild, V. Sguera, Gaurav Jaiswal, and S.
Tsygankov, for their comments and suggestions that allowed us to
improve the clarity of the original version. Special thanks to
Nicola Masetti  for comparing our data with the data in his web page
http://www.iasfbo.inaf.it/~masetti/IGR/sources/17391.html. A. Taani gratefully acknowledges support and
hospitality from the Institute of High Energy Physics, Chinese
Academy of Sciences through the CAS President's International
Fellowship Initiative (PIFI).

\section*{DATA AVAILABILITY}
The data that support the findings of this study are available
from the corresponding author upon reasonable request.

\end{document}